\documentclass{article}
\usepackage{spconf,amsmath,graphicx}
\usepackage{url}
\usepackage{graphicx}
\usepackage{subfigure}
\usepackage{cite}
\usepackage{tablefootnote}
\usepackage{booktabs}
\usepackage{tablefootnote}
\usepackage{multirow}
\usepackage{xcolor}

\title{Genre-conditioned Acoustic Models for Automatic Lyrics Transcription of Polyphonic Music}

\name{Xiaoxue Gao$^1$, Chitralekha Gupta$^2$ and Haizhou Li$^3$}
\address{$^1$Department of Electrical and Computer Engineering, National University of Singapore, Singapore\\
$^2$Department of Communications and New Media, National University of Singapore, Singapore\\
$^3$The Chinese University of Hong Kong, Shenzhen, China\\}

\begin{document}
\ninept
\maketitle
\vspace{-0.1cm}
\begin{abstract}
\vspace{-0.05cm}
Lyrics transcription of polyphonic music is challenging not only because the singing vocals are corrupted by the background music, but also because the background music and the singing style vary across music genres, such as pop, metal, and hip hop, which affects lyrics intelligibility of the song in different ways. In this work, we propose to transcribe the lyrics of polyphonic music using a novel genre-conditioned network. The proposed network adopts pre-trained model parameters, and incorporates the genre adapters between layers to capture different genre peculiarities for lyrics-genre pairs, thereby only requiring lightweight genre-specific parameters for training. Our experiments show that the proposed genre-conditioned network outperforms the existing lyrics transcription systems. 

\end{abstract}
\vspace{-0.3cm}
\section{Introduction}
\label{sec:intro}
\vspace{-0.2cm}
Despite much progress in automatic speech recognition (ASR)~\cite{povey2016,sainath2013deep} and deep learning~\cite{gao2019speaker,gao2020personalized,gao2018nus}, there are fewer studies in lyrics transcription of polyphonic music. The aim of automatic lyrics transcription is to transcribe lyrics from a song that contains singing vocals mixed with background music. Lyrics transcription has attracted a lot of interest to aid applications such as automatic generation of karaoke lyrical content, music video subtitling and query-by-singing \cite{mesaros2013singing}. 

Background music often correlates with the singing vocals, making the task of lyrics transcription challenging. Past studies have broadly taken two approaches to deal with the background music: 1) an \textit{extraction-transcription} approach, in which singing vocals are extracted from the polyphonic music \cite{gupta2019,mesaros2010automatic,dzhambazov2015modeling} as a pre-processing step, and lyrics are transcribed from these extracted vocals; and 2) a \textit{music-aware} approach, whereby the background music knowledge is used to help the transcription model \cite{stoller2019,gupta2019automatic}.

In the \textit{extraction-transcription} approach, several singing vocal separation techniques have been studied to suppress the background music and extract the singing vocals from the polyphonic music that are used for acoustic model training \cite{gupta2019,mesaros2010automatic,dzhambazov2015modeling,fujihara2011lyricsynchronizer}. 
However, due to imperfections in music removal and the distortions associated with the inversion of a magnitude spectral representation, the extracted time-domain singing vocal signals often contain artifacts. Acoustic model trained on such extracted vocals are far from perfection~\cite{gupta2019,gupta2019automatic}. 

Another way is to use acoustic model trained on clean singing vocals, and at the time of inference, apply source separation technique to extract singing vocal from the input polyphonic song to transcribe lyrics~\cite{guptalyrics,dabikesheffield,demirel2021low}. However, this two-step approach not only needs a large amount of data for training the vocal separation model as well as the acoustic model~\cite{demirel2020automatic}, but also suffers from mismatch between the acoustic features between training and testing, thereby causing degradation of the performance of acoustic modeling in lyrics recognition. 

Gupta et al.~\cite{gupta2019acoustic} found that an acoustic model trained on a large amount of solo singing data, when adapted with a small amount of in-domain polyphonic data outperforms the lyrics transcription performance of solo-singing acoustic models adapted with extracted singing vocals. This suggests that the polyphonic data, i.e.~singing vocals+background music, helps in learning the spectro-temporal variations of the background music more than the extracted vocals.

In the \textit{music-aware} approach, instead of removing the background music, the systems~\cite{stoller2019,gupta2019automatic} make use of music information by directly training with the polyphonic music input for lyrics recognition. For example, Stoller et al.~\cite{stoller2019} adopted an end-to-end wave-U-net model to predict character probabilities from the polyphonic audio, while Gupta et al.~\cite{gupta2019automatic} designed a music genre-based kaldi-based acoustic model, that outperforms all previous techniques. These studies show that the task of lyrics transcription in polyphonic music can benefit from the knowledge of background music. 

Different genres exhibit significantly different levels of lyric intelligibility in polyphonic music~\cite{condit2015catching}, since the genres vary in their musical characteristics such as instrumental accompaniment, singing vocal loudness, syllable rate, reverberation, and singing style~\cite{condit2015catching,gupta2019automatic}. In \cite{gupta2019automatic}, the music genre tag was embedded in the pronunciation dictionary at the time of training, while at the time of inference, genre information was not provided. In contrast, in this work, we believe that the predictable genre-class information would help an automated lyrics transcription system with lyrics intelligibility. We propose a music genre-conditioned training strategy to adapt an end-to-end lyrics transcription system according to the music genre. 
Inspired by the success of adaptive fine-tuning with pre-trained models in natural language processing~\cite{houlsby2019parameter} and speech translation~\cite{le2021lightweight,li2020multilingual,xu2021stacked}, we propose to incorporate genre-specific adapters to a pre-trained transformer-based polyphonic lyrics transcription model~\cite{gao2021tran}.

\vspace{-0.2cm}
\section{Genre-conditioned automatic lyrics transcription}
\vspace{-0.2cm}

Music genre are categorical labels created by human experts to characterize and structure pieces of music~\cite{genussov2010musical,tzanetakis2002musical}. Musical genres differ from each other in their instrumental accompaniment, rhythmic structure, vocal harmonization, reverberation and pitch content of the music~\cite{genussov2010musical,tzanetakis2002musical}. Lyrics intelligibility in polyphonic music is found to be influenced by these genre specific characteristics~\cite{gupta2019automatic,condit2015catching}. Factors such as instrumental accompaniment, vocal harmonization, and reverberation are expected to interfere with lyric intelligibility, while predictable rhyme schemes and semantic context might improve intelligibility \cite{condit2015catching}. For example, as observed in \cite{condit2015catching}, in \textit{metal} songs, the accompaniment is loud and interferes with the vocals, while is relatively softer in \textit{jazz}, \textit{country}, and \textit{pop} songs. As a result ``Death Metal'' songs shows lyrics intelligibility scores of zero, while ``Pop'' songs achieve scores close to 100\% ~\cite{condit2015catching}. Another difference is the syllable rate between genres. In \cite{condit2015catching}, it was observed that \textit{rap} songs, that have a higher syllable rate, show lower lyric intelligibility than other genres. We expect that these factors are important for building an automatic lyrics transcription framework.
\vspace{-0.1cm}
\subsection{Genre-conditioned Acoustic Model}
\vspace{-0.2cm}

\begin{figure}[t]
\vspace{-0.4cm}
\flushleft 
\includegraphics[width=87mm]{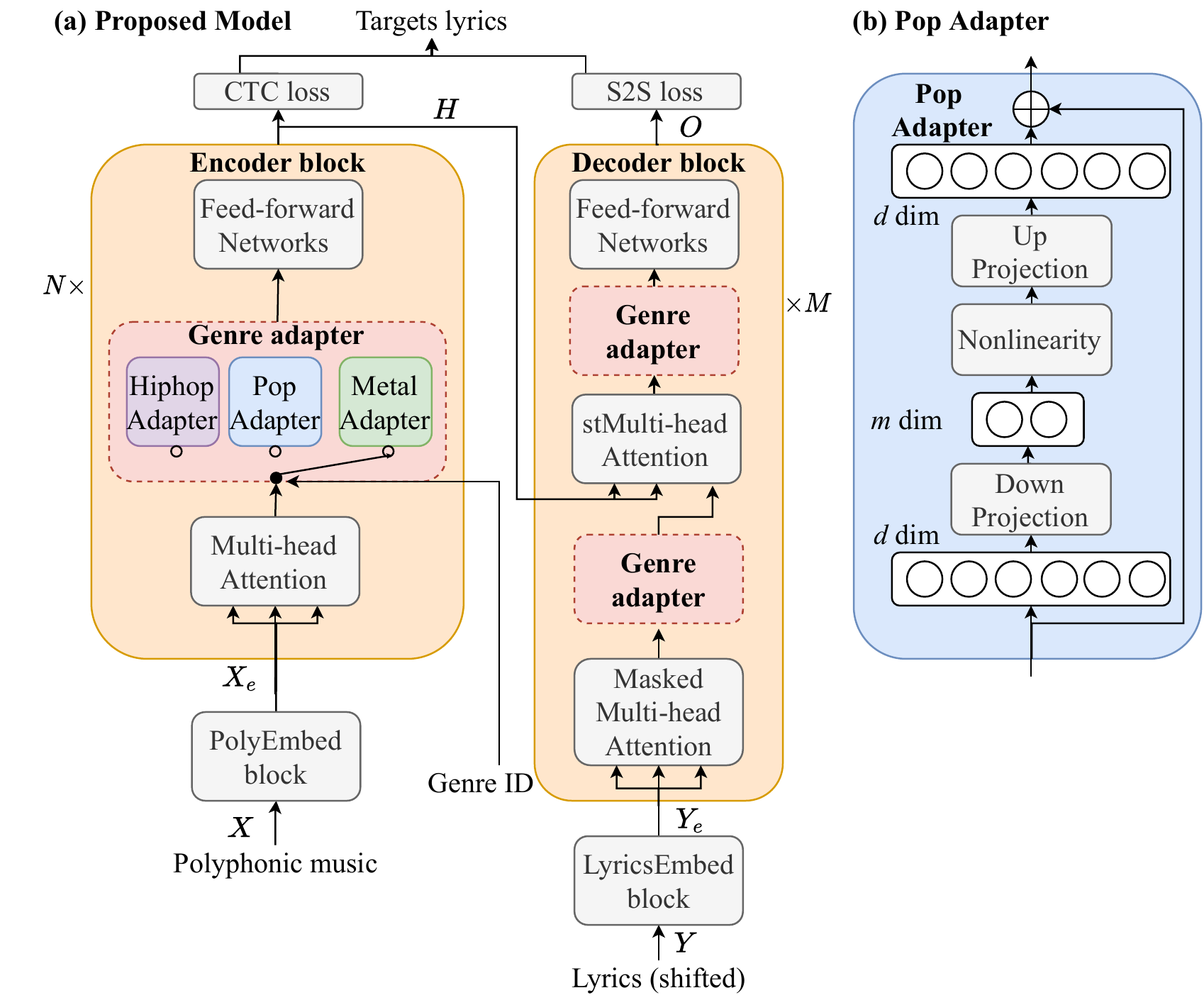}
\vspace{-0.6cm}
\caption{An overview architectures of (a) the proposed genre-conditioned lyric transcriber with a Transformer architecture. For simplicity, layer normalization and residual connection are omitted, and genre ID is applied for all genre adapters; and (b) the pop adapter, which has the same design for metal and hiphop adapters.}
\vspace{-0.6cm}
\label{base-overview}
\end{figure}

The proposed lyric transcriber consists of an encoder block, and a decoder block. The encoder block is designed to produce an acoustic latent representation from the input polyphonic music, while the decoder block combines the encoded acoustic representation and the previously predicted lyrical information to transcribe the lyrics. 
The focus of this study is a simple yet effective genre-specific adapter and its integration with a pre-trained model. 

Our framework is based on transformer~\cite{vaswani2017attention} with joint encoder-decoder and connectionist temporal classification (CTC) architecture~\cite{nakatani2019improving}. We incorporate a genre-conditioned encoder that converts the input polyphonic acoustic features to intermediate genre-specific representations, and a genre-conditioned decoder that predicts lyrical tokens, i.e. sub-words in this paper, one at a time given the intermediate genre-specific representations and the previously predicted lyrical tokens in an auto-regressive manner, as illustrated in Fig~\ref{base-overview} (a).
The core module of the genre-conditioned training framework is a genre adapter, that employs a down-projection and an up-projection with non-linear activation function to make use of different genre temporal context of the polyphonic music input sequence. 
\vspace{-0.2cm}
\subsubsection{Genre Broadclasses}
\vspace{-0.1cm}
Music has been divided into different genres in many different and overlapping ways, based on a shared set of characteristics \cite{tzanetakis2002musical}. The main difference between genres that affects lyrics intelligibility are the relative singing vocal volume compared with background music and syllable rate~\cite{condit2015catching,gupta2019automatic}. Based on the shared characteristics between genres that affect lyrics intelligibility, Gupta et al.~\cite{gupta2019automatic} mapped all genres into three broad genre classes namely \textit{pop}, \textit{metal} and \textit{hiphop}. The song containing some rap along with electronic music were mapped to \textit{hiphop} broadclass, that includes genres such as Rap, Hip Hop, and Rhythms \& Blues. Songs with loud and dense background music were categorized as \textit{metal}, that includes genres such as Metal and Hard Rock. And songs with clear and louder vocals under genres Pop, Country, Jazz, Reggae etc. were categorized as \textit{pop} broadclass. In this work, we follow the same categorization and broadclasses.

\vspace{-0.3cm}
\subsubsection{Genre-specific Adapters}
\vspace{-0.2cm}
We incorporate a genre-adapter as a new block of layers between the transformer layers of a pre-trained base model~\cite{gao2021tran} (Fig.~\ref{base-overview} (a)). A genre-specific adapter first projects the original \textit{d}-dimensional input feature into a smaller bottleneck dimension, \textit{m}, applies a non-linearity, and then projects back to \textit{d} dimension. We expect that this bottleneck lower-dimensional feature, \textit{m}, captures genre-specific information. The genre-specific adapter itself has a skip-connection internally as shown in Fig.~\ref{base-overview} (b). With the skip-connection, if the parameters of the projection layers are initialized to near-zero, the module is initialized to an approximate identity function.

The idea is that the holistic polyphonic representation, as captured by the pre-trained base model (which is trained on polyphonic, i.e.~singing+background music data), provides information that is common across the different genres, while the genre adapters capture the genre-specific characteristics, thereby tuning the base model as per genre identity of the song. The genre broadclass ID, for example pop, is given to the genre adapter to choose the corresponding adapter pathway, i.e.~pop adapter, to characterize the corresponding genre attributes as shown in Fig.~\ref{base-overview} (b). Most of the weights of the pre-trained base model are untouched, except the layer normalization~\cite{ba2016layer} and the source-target multi-head attention (stMulti-head Attention in Fig.~\ref{base-overview} (a)). 

In this work, we explore inserting the genre-adapter at multiple locations in the pre-trained base model architecture: in the encoder, and at two places in the decoder. The genre-conditioned encoder would adapt the network according to the genre-specific acoustic characteristics, while the genre-conditioned decoder would also capture the lyrical characteristics specific to genre. In the following sub-sections, we describe the genre-conditioned modules individually.
\vspace{-0.2cm}
\subsubsection{Genre-conditioned Encoder}
\label{Poly-Transformer Encoder}
\vspace{-0.1cm}
The genre-conditioned encoder consists of an embedding block (PolyEmbed) and $N$ identical encoder blocks where each encoder block contains a multi-head attention (MHA)~\cite{vaswani2017attention}, a genre adapter, and a position-wise feed-forward network (FFN). The input sequence $\mathbf{X}$ is first encoded into $\mathbf{X}_{e}$ by PolyEmbed block using subsampling and positional encoding (PE)~\cite{vaswani2017attention}. The genre-conditioned encoder then transform $\mathbf{X}_{e}$ into a hidden representation $\mathbf{H}$ via the MHA, the genre adapter and the FFN. Residual connection~\cite{he2016deep} and layer normalization~\cite{ba2016layer} are employed inside each of the encoder blocks for MHA and FFN as in~\cite{vaswani2017attention}. 
\vspace{-0.2cm}
\begin{equation}
\centering
\vspace{-0.3cm}
\begin{split}     
       \mathbf{X}_{e} =  \text{PolyEmbed}(\mathbf{X}),\\
       \mathbf{H} = \text{GenreEncoder}(\mathbf{X}_{e})
\end{split}
\vspace{-0.2cm}
\end{equation}
\vspace{-0.5cm}

\subsubsection{Genre-conditioned Decoder}
The genre-conditioned decoder consists of a textual embedding block (LyricsEmbed) and $M$ identical decoder blocks, where each decoder block has a masked MHA, a MHA, a FFN and genre adapters.  
During training, $\mathbf{Y}$ represents the lyrical token history that is offset right by one position, however, at run-time inference, it represents the previous predicted token history. $\mathbf{Y}$ is first converted to lyrics token embedding $\mathbf{Y}_{e}$ via a LyricsEmbed block, that consists of an embedding layer, and a positional encoding (PE) operation.
\vspace{-0.1cm}
\begin{equation}
\vspace{-0.1cm}
\centering
\begin{split}    
       \mathbf{Y}_{e} =  \text{LyricsEmbed}(\mathbf{Y}),\\
       \mathbf{O} = \text{GenreDecoder}(\mathbf{H},\mathbf{Y}_{e})
\end{split}
\end{equation}

The lyrics embedding $\textbf{Y}_{e}$ is fed into the masked MHA that ensures causality, i.e.~the predictions for current position only depends on the past positions. The output of the masked MHA is fed into the genre-adapter, whose outputs and the acoustic encoding $\textbf{H}$ are then fed to the source-target MHA for capturing the relationship between acoustic information $\textbf{H}$ and the genre-informed textual information. The output of the MHA is then passing through the another genre adapter and a FFN to generate the decoder output $\textbf{O}$. The residual connection~\cite{he2016deep} and layer normalization~\cite{ba2016layer} are also employed in each of the decoder blocks for masked MHA, stMHA and FFN~\cite{vaswani2017attention}.

\begin{table}[t]
\vspace{-1.1cm}
\centering
\caption{A description of polyphonic music dataset that consists of DALI and NUS collections.}
\label{tab:datasets}
\small
\begin{tabular}{l|l|rrr}
\toprule
\multicolumn{2}{c}{ }      & \textbf{\# songs} & \textbf{\# lines} & \textbf{duration} \\ \midrule
\multirow{2}{*}{Poly-train} & DALI-train & 3,913          & 180,034                & 208.6 hours       \\ 
                            & NUS        & 517            & 264,62                 & 27.0 hours        \\ \midrule
\multirow{2}{*}{Poly-dev}   & DALI-dev   & 100            & 5,356                  & 3.9 hours         \\ 
                            & NUS        & 70             & 2,220                  & 3.5 hours         \\ \midrule
\multirow{3}{*}{Poly-test}  & Hansen     & 10             & 212                    & 0.5 hour          \\ 
                            & Jamendo    & 20             & 374                    & 0.9 hour          \\ 
                            & Mauch      & 20             & 442                    & 1.0 hour          \\ \bottomrule
\end{tabular}
\vspace{-0.6cm}
\end{table}
\vspace{-0.2cm}
\subsubsection{Multi-task Learning Objective}
We employ both CTC~\cite{nakatani2019improving} and sequence-to-sequence (S2S) objective functions for model training. 
The $M$ decoder blocks are followed by a linear projection and softmax layers, that converts the decoder output $\mathbf{O}$ into a posterior probability distribution of the predicted lyrical token sequence $\textbf{G}_{s2s}$. The S2S loss, $\mathcal{L}^{\text{S2S}}$, is the cross-entropy of the ground-truth lyrical token $\mathbf{R}$ and $\mathbf{G}_{s2s}$. Also, a linear transform is applied on $\textbf{H}$ to obtain the token posterior distribution $\textbf{G}_{ctc}$. CTC loss is computed between $\textbf{G}_{ctc}$ and $\textbf{R}$. The network is trained to minimize both S2S and CTC losses jointly with an objective function $\mathcal{L}_{\text{Genre-con}}$,

\vspace{-0.29cm}
\begin{equation}
\vspace{-0.29cm}
\begin{split} 
 \mathcal{L}_{\text{Genre-con}} = \alpha \mathcal{L}^{\text{CTC}} + (1-\alpha) \mathcal{L}^{\text{S2S}},\\
  \mathcal{L}^{\text{CTC}} = \text{Loss}_{\text{CTC}}(\mathbf{G}_{ctc},\mathbf{R}),\\
  \mathcal{L}^{\text{S2S}} = \text{Loss}_{\text{S2S}}(\mathbf{G}_{s2s},\mathbf{R})   
\end{split}
\label{polyLoss}
\vspace{-0.26cm}
\end{equation}

where $\alpha \in [0,1]$, $\mathbf{R}$ is the ground-truth lyrical token sequence. 
During run-time inference, the genre-conditioned model converts input polyphonic acoustic features to output lyrical token sequence.
\vspace{-0.2cm}
\section{Experiments}
\label{Experiments}
\vspace{-0.3cm}
\subsection{Datasets}
\label{sec:datasets}
As shown in Table \ref{tab:datasets}, the polyphonic music training dataset, Poly-train, consists of the DALI-train~\cite{meseguer2018dali} dataset and a NUS proprietary collection. The DALI-train dataset consists of 3,913 English polyphonic song tracks \footnote{There are a total of 5,358 audio tracks in DALI, but we only have access to  3,913 English audio links.}. The dataset is processed into 180,034 lyrics-transcribed audio lines with a total duration of 208.6 hours. The NUS collection dataset consists of 517 popular English songs. We obtain its line-level lyrics boundaries using the state-of-the-art audio-to-lyrics alignment system~\cite{9054567}, leading to 26,462 lyrics-transcribed audio lines with a total duration of 27.0 hours.

The Poly-dev dataset consists of the DALI-dev dataset of 100 songs from DALI dataset \cite{9054567}, and 70 songs from a NUS proprietary collection. We adopt three widely used test sets -- Hansen\cite{hansen2012recognition},  Jamendo\cite{stoller2019}, and Mauch\cite{mauch2010lyrics} to form the Poly-test as shown in Table~\ref{tab:datasets}. The test datasets are English polyphonic songs, that are manually segmented into line-level segments of average duration 8 seconds. We transcribe the lyrics line-by-line, instead of the whole song, to avoid possible accumulated errors in the Viterbi decoding that occur in long duration audio clips ~\cite{moreno1998recursive,gupta2018semi}.

Genre tags for all of the NUS collection dataset and most of the songs in DALI-train and DALI-dev sets are provided in their metadata, except for 840 songs in DALI. For these remaining songs, we apply an automatic genre recognition implementation~\cite{genre2019web}, that has 80\% accuracy, to obtain their genre IDs. For the test sets, we scan the web to find the genre tags of each song. For each song in train, dev, and test sets, the genre tags are mapped to one of the three genre broadclasses, pop, metal, or hiphop, according to the mapping described in~\cite{gupta2019automatic}. The statistics of music genre distribution for the polyphonic music datasets are provided in Table~\ref{stats}.

\begin{table}[t]
\centering
\vspace{-1.1cm}
\caption{The genre distribution for polyphonic music Dataset.}
\small
\begin{tabular}{l|ccc}
\toprule
    \textbf{Statistics}  & \textbf{Metal}  & \textbf{Pop}   & \textbf{Hiphop} \\ \midrule
      Percentage in Poly-train & 35\% & 59\% & 6\% \\ 
            Percentage in Poly-dev & 48\% & 49\% & 3\% \\ 
            Percentage in Poly-test & 34\% & 56\% & 10\% \\ 
\bottomrule
\end{tabular}
\label{stats}
\vspace{-0.6cm}
\end{table}

\vspace{-0.3cm}
\subsection{Experimental Setup}
\vspace{-0.1cm}
We use ESPnet~\cite{watanabe2018espnet} with pytorch backend to build all the models (including the pre-trained base model~\cite{gao2021tran} and the proposed genre-conditioned models).
We extract 83-dimensional filterbank features along with pitch from the audio files with a window of 25 ms, and a frame-shift of 10 ms.
We use sub-words as the model token units for the task of lyrics transcription, and 5,000 sub-words are generated using byte-pair encoding (BPE).
All models are trained with the Adam optimizer with a Noam learning rate decay, 25,000 warmup steps, 5,000,000 batch-bin, and 100 epochs as in~\cite{vaswani2017attention}. The PolyEmbed block contains two CNN blocks with a kernel size of 3 and a stride size of 2. 

Other parameters of all the models follow the default settings in the published LibriSpeech model (LS)~\footnote{Pretrained LS model ``pytorch large Transformer with specaug (4 GPUs) + Large LSTM LM" from the ESPNET github \url{https://github.com/espnet/espnet/blob/master/egs/librispeech/asr1/RESULTS.md}.}, where attention dim is 512, the number of heads is 8 in MHA, FFN laryer dim is 2048, the interpolation factor $\alpha$ between CTC loss and S2S loss (Eq.~\ref{polyLoss}) is set to 0.3, and there are 12 encoder blocks and 6 decoder blocks ($N=12$ and $M=6$). We follow the default setting in ESPnet~\cite{watanabe2018espnet} to average the best 5 validated model checkpoints on the development set Poly-dev (Table~\ref{tab:datasets}) to obtain the final model. 
We follow the common joint decoding approach~\cite{hori2018end,nakatani2019improving}, which takes CTC prediction into account during decoding where we set the same decoding parameters (penalty, beam width and CTC decoding weight are set to 0.0, 10 and 0.3, respectively). 

The pre-trained base model~\cite{gao2021tran} is trained using Poly-train, and Poly-dev as the development set and Poly-test to evaluate, and the base model is pre-trained by solo-singing model using the English solo-singing dataset $\textit{Sing! 300} \times \textit{30} \times \textit{2}$~\footnote{The audio files can be accessed from https://ccrma.stanford.edu/damp/}, whose weights are initialized by the published LS model. In the genre-conditioned models, we insert the genre adapters in the pre-trained base model, as shown in Figure \ref{base-overview} (a). In each genre-adapter, the down-projection and up-projection layers are linear layers with $d=512$ and $m=256$, along with a ReLU non-linearity function.

\vspace{-0.9cm}
\section{Results and Discussion}
\label{Results}
\vspace{-0.3cm}
We study the effects of genre-conditioned approach and the places to plug-in the genre adapters. We also compare the performance of the proposed models with the SOTA systems for lyrics transcription of polyphonic music and conducts an ablation study. We report the performance in terms of word error rate (WER), which is the ratio of the total number of insertions, substitutions, and deletions with respect to the total number of words per song. 
\vspace{-0.4cm}
\subsection{Genre-Conditioned Training}
\vspace{-0.2cm}
We study the effectiveness of genre-conditioned training approach compared to the pre-trained base model. The lyrics transcription performance of the framework with and without the genre-conditioned training is presented in Table~\ref{genre_results}. In Genre MHA+MaskMHA model, the genre adapter is inserted in the encoder and the two places in the decoder, as shown in Fig.~\ref{base-overview}. In Genre MHA model, we insert the genre-adapter in the encoder and only at one place in the decoder, i.e.~the one after stMHA. 

The two genre-conditioned models (Genre MHA and Genre MHA+MaskMHA) outperform the base model for pop and metal songs, and comparable with the base model for the hiphop songs (only 5 hiphop songs in Poly-test). This suggests that the genre-specific adapters are able to capture the differences across music genres in a rich polyphonic environment. One should note that the number of pop and metal songs in the train and dev datasets are considerably more than that of hiphop songs (only 6\% in Poly-train and 3\% in Poly-dev are hiphop songs), as illustrated in Table~\ref{stats}. The lack of data for training the hiphop adapter results in sub-par performance of the system on the songs of this genre. 

We further investigate the effect of different places of inserting the genre adapters. The genre-adapter in the encoder captures the genre-specific information from the acoustic features. The first adapter in the decoder intends to capture genre-related features from the previously predicted lyrics, while the second adapter in the decoder captures the genre-information from the combination of acoustic and lyrical features. We find that the Genre MHA model (that does not have the first decoder adapter which only takes lyrical features as input), performs better than the Genre MHA+MaskMHA model (that has all the three adapters) for metal and pop genres. This observation aligns with the fact that the characteristics that define these music genres depend more on the acoustic features (such as loudness, musical instrument types), than on textual features. For hiphop, however, the textual features can indicate the high syllable rate of rap songs, thereby, the first adapter of the decoder might help, resulting in a slightly better performance of Genre MHA+MaskMHA for hiphop genre. However, as mentioned earlier, the amount of training, dev, and test data for hiphop genre was small. Therefore, hiphop genre needs further investigation in the future.

For ablation study purpose, the Genre MHA Ablation experiment is conducted with one common adapter with pop, hiphop and metal parameters shared. The proposed Genre MHA outperforms the Genre MHA Ablation for pop and hiphop songs. The genre-specific adapter is especially beneficial to the hiphop songs compared with the common adapter, which suggests that the different properties of different genres should be considered separately.

\vspace{-0.3cm}
\subsection{Comparison with Prior Studies}
\vspace{-0.1cm}
We compare the proposed models with the existing approaches~\cite{stoller2019,9054567,dabikesheffield,demirel2020recursive,gaolyrics,gao2021tran} for lyrics transcription of polyphonic music in Table~\ref{genre_results}. Stoller et al.'s \cite{stoller2019} system is based on E2E Wave-U-Net framework, while the remaining systems~\cite{9054567,dabikesheffield,demirel2020recursive,gaolyrics} are based on the Kaldi. A subset of these systems~\cite{dabikesheffield,demirel2020recursive,gaolyrics} were submitted to the lyrics transcription task in MIREX 2020, where the GGL1 system~\cite{gaolyrics} outperformed other submissions\footnote{\url{https://www.music-ir.org/mirex/wiki/2020:Lyrics_Transcription_Results}}.  

We first report the lyrics transcription performance of all existing systems on the same test sets for whole songs evaluation. As can be seen, GGL1 \cite{gaolyrics} performs the best among all and considered as the published state-of-the-art system. Therefore, we use GGL1~\footnote{The results of GGL1 and GGL2 are obtained by standard Kaldi recipe with scoring, which are slightly different from those at MIREX2020 website.} as our point of comparison for the genre-based analysis next.
To avoid accumulated errors while decoding long duration songs, we segment the songs in the test sets into lines, as described in Section \ref{sec:datasets}. We further report the lyrics transcription results for these segmented testsets across different genres. We observe that the base model performs better than GGL1 for all the genres, implying that the end-to-end transformer based framework works better than the kaldi-based conventional framework. Our proposed models outperform the base model for pop and metal songs, which indicates the superiority of genre-conditioned training over the base model. The genre models and the base model also outperform GGL1 across all the test data with relative 8\%-19\% improvements. This indicates the superiority of genre-conditioned training and the end-to-end transformer-based model over the conventional multi-step ASR pipeline.

\vspace{-0.9cm}
\begin{table}[t]
\vspace{-0.3cm}
\centering
\caption{Comparison between the proposed genre-adapter solutions and other existing competitive solutions to lyrics transcription (WER\%) of polyphonic music.}
\begin{tabular}{l|ccc}
\toprule
\textbf{Whole songs test}         & \textbf{Hansen} & \textbf{Jamendo} & \textbf{Mauch} \\ \midrule
DS~\cite{stoller2019}                          & -               & 77.80            & 70.90          \\ 
RB1~\cite{dabikesheffield}                         & 83.43           & 86.70            & 84.98          \\ 
DDA2~\cite{demirel2020recursive}                         & 74.81           & 72.15            & 75.39          \\ 
DDA3~\cite{demirel2020recursive}                         & 77.36           & 73.09            & 80.66          \\ 
CG~\cite{9054567}                          & -               & 59.60            & 44.00          \\ 
GGL2~\cite{gaolyrics}                         & 48.11           & 61.22            & 45.35          \\ 
GGL1~\cite{gaolyrics}                     & 45.87           & 56.76            & 43.76          \\ 
\toprule
  \textbf{Line-level test}    & \textbf{Metal}  & \textbf{Pop}   & \textbf{Hiphop} \\ \midrule
  GGL1~\cite{gaolyrics} & 59.70& 37.07 &57.08  \\ 
Base model~\cite{gao2021tran} & 50.04&36.52  &\textbf{51.19}  \\ 
Genre MHA &  \textbf{48.17}  &  \textbf{33.34} & 52.32 \\
Genre MHA Ablation &48.05    &33.41   &55.42\\
Genre MHA+MaskMHA & 48.22 & 33.86& 51.55 \\ \bottomrule

\end{tabular}
\vspace{-0.6cm}
\label{genre_results}
\end{table}

\vspace{+0.3cm}
\section{Conclusion}
\label{Conclusion}
\vspace{-0.3cm}
We propose a novel genre-conditioned lyrics transcription network architecture that captures genre-specific information from the acoustics and the lyrics, and adapts the network as per the music genre. The proposed approach injects genre-specific adapter into the backbone transformer pre-trained model to interpret different genre attributes in a single network. We have presented a study of adapters for lyrics-genre pairs in polyphonic music, and shown the genre adapters can provide genre-related knowledge to help with music interference problem. Integrating genre-adapters with existing pre-trained model also shows the flexibility of using these adapters to explore different kinds of music data for the development of lyrics transcription system for polyphonic music.

\footnotesize
\textbf{Acknowledgement}
This research is supported by the Agency for Science, Technology and Research (A*STAR) under its AME Programmatic Funding Scheme (Project No. A18A2b0046). This work is supported by A*STAR under its RIE2020 Advanced Manufacturing and Engineering Domain (AME) Programmatic Grant (Grant No. A1687b0033, Project Title: Spiking Neural Networks).

\footnotesize
\bibliographystyle{IEEEtran}
\bibliography{strings}

\end{document}